\documentclass{iaus}
\usepackage{graphicx}

\title[Measuring interstellar magnetic fields] 
{Measuring interstellar magnetic fields by radio synchrotron
emission}

\author[Rainer Beck]   
{Rainer Beck}

\affiliation{Max-Planck-Institut f\"ur Radioastronomie, Auf dem
H\"ugel 69, 53121 Bonn, Germany \break email:
rbeck@mpifr-bonn.mpg.de}

\pubyear{2009}
\volume{259}  
 \jname{Cosmic Magnetic Fields: From Planets, to
Stars and Galaxies} \editors{K.G. Strassmeier, A.G. Kosovichev \&
J.E. Beckman, eds.}

\begin{document}

\maketitle

\begin{abstract}
Radio synchrotron emission, its polarization and its Faraday
rotation are powerful tools to study the strength and structure of
interstellar magnetic fields. The total intensity traces the
strength and distribution of total magnetic fields. Total fields in
gas-rich spiral arms and bars of nearby galaxies have strengths of
20--30~$\mu$Gauss, due to the amplification of turbulent fields, and
are dynamically important. In the Milky Way, the total field
strength is about $6~\mu$G near the Sun and several $100~\mu$G in
filaments near the Galactic Center. -- The polarized intensity
measures ordered fields with a preferred orientation, which can be
regular or anisotropic fields. Ordered fields with spiral structure
exist in grand-design, barred, flocculent and even in irregular
galaxies. The strongest ordered fields are found in interarm
regions, sometimes forming ``magnetic spiral arms'' between the
optical arms. Halo fields are X-shaped, probably due to outflows. --
The Faraday rotation of the polarization vectors traces coherent
regular fields which have a preferred direction. In some galaxies
Faraday rotation reveals large-scale patterns which are signatures
of dynamo fields. However, in most galaxies the field has a
complicated structure and interacts with local gas flows. In the
Milky Way, diffuse polarized radio emission and Faraday rotation of
the polarized emission from pulsars and background sources show many
small-scale and large-scale magnetic features, but the overall field
structure in our Galaxy is still under debate.

\end{abstract}

\firstsection 

\section{Introduction}

Interstellar magnetic fields were discovered already in 1932 by Karl
Guthe Jansky who first detected diffuse low-frequency radio emission
from the Milky Way, but the explanation as synchrotron emission was
given only in 1950 by Karl Otto Kiepenheuer. The sensitivity of
radio observations has improved by several orders of magnitude in
the past decades, and synchrotron emission was detected from the
interstellar medium (ISM) in almost all star-forming galaxies, in
galaxy halos and the intracluster medium, proving that a large
fraction of the Universe is permeated by magnetic fields. However,
in spite of our increasing knowledge on interstellar magnetic
fields, many important questions are unanswered, especially their
first occurrence in young galaxies, their amplification when
galaxies evolved, and their effect on galaxy dynamics.

As magnetic fields need illumination by cosmic-ray electrons to
become observable by synchrotron emission, which are generated in
star-forming regions or intracluster shocks, we do not know yet
whether magnetic fields also exist in radio-quiet elliptical or
dwarf galaxies or in the general intergalactic medium (IGM).
Progress can be expected from using Faraday rotation which does not
need cosmic rays, only magnetic fields and thin ionized gas. One of
the research areas of the forthcoming radio telescopes (LOFAR,
ASKAP, SKA) will be the search for Faraday rotation in these objects
against polarized background sources (Gaensler, this volume). The
SKA will also be needed to detect magnetic fields in young galaxies
(\cite{beck04,arshakian08}).

\section{Tools to measure interstellar magnetic fields}

Most of what we know about galactic and intergalactic magnetic
fields comes through the detection of radio waves. {\em Zeeman
splitting}\ of radio spectral lines is the best method to directly
measure the field strength (Heiles, this volume).

\begin{figure}
\begin{minipage}[t]{6.7cm}
\vspace{0.3cm}
\includegraphics[bb = 32 99 537 702,width=6.7cm,clip=]{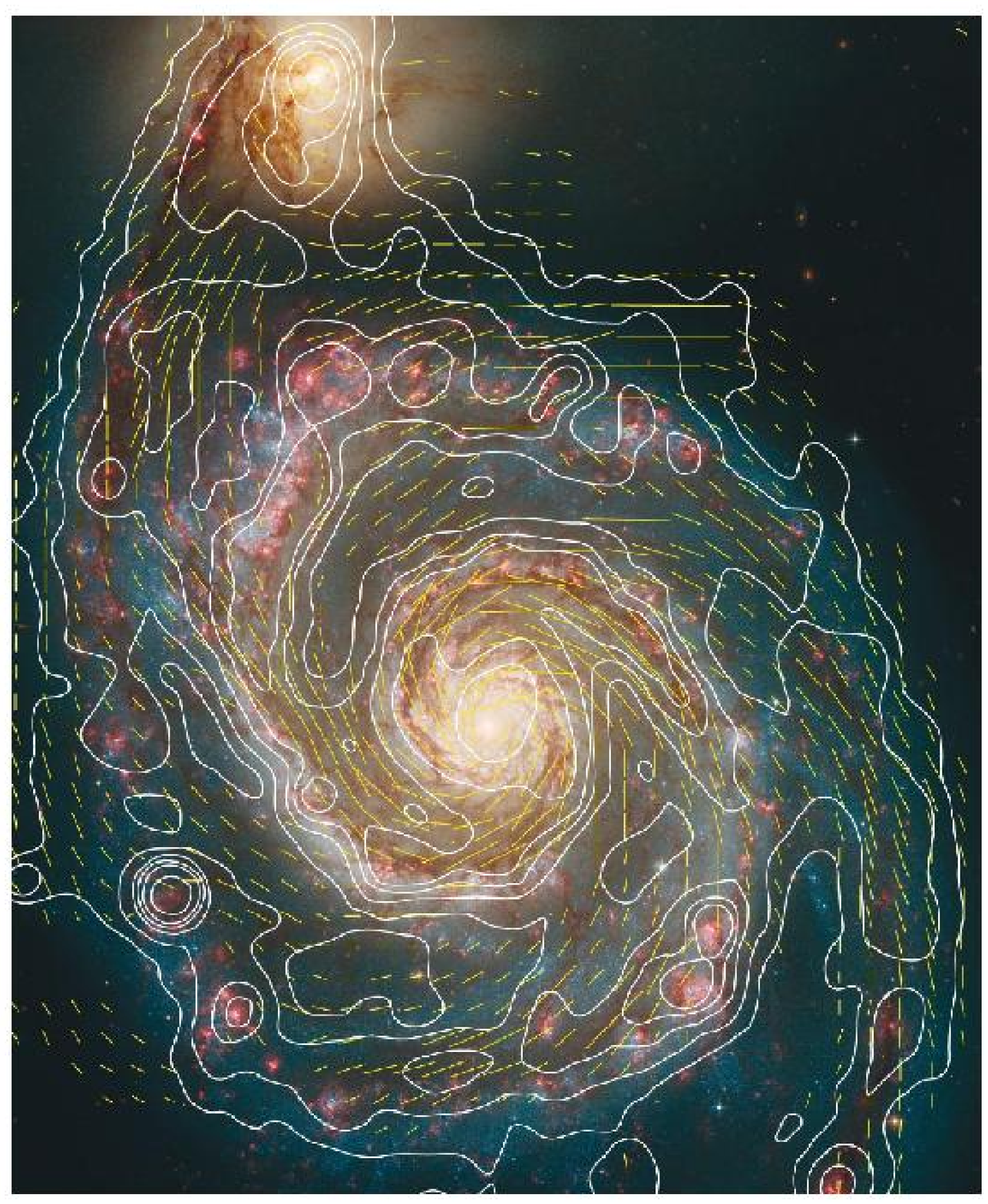}
\caption{Total radio emission (contours) and $B$--vectors of M~51,
combined from observations at 6~cm wavelength with the VLA and
Effelsberg telescopes and smoothed to 15'' resolution (Fletcher et
al., in prep.), overlaid onto an optical image from the HST
(Copyright: MPIfR Bonn and \textit{Hubble Heritage Team}. Graphics:
\textit{Sterne und Weltraum}).} \label{fig:m51}
\end{minipage}\hfill
\begin{minipage}[t]{6.5cm}
\vspace{0.3cm}
\includegraphics[bb = 47 85 522 702,width=6.4cm,clip=]{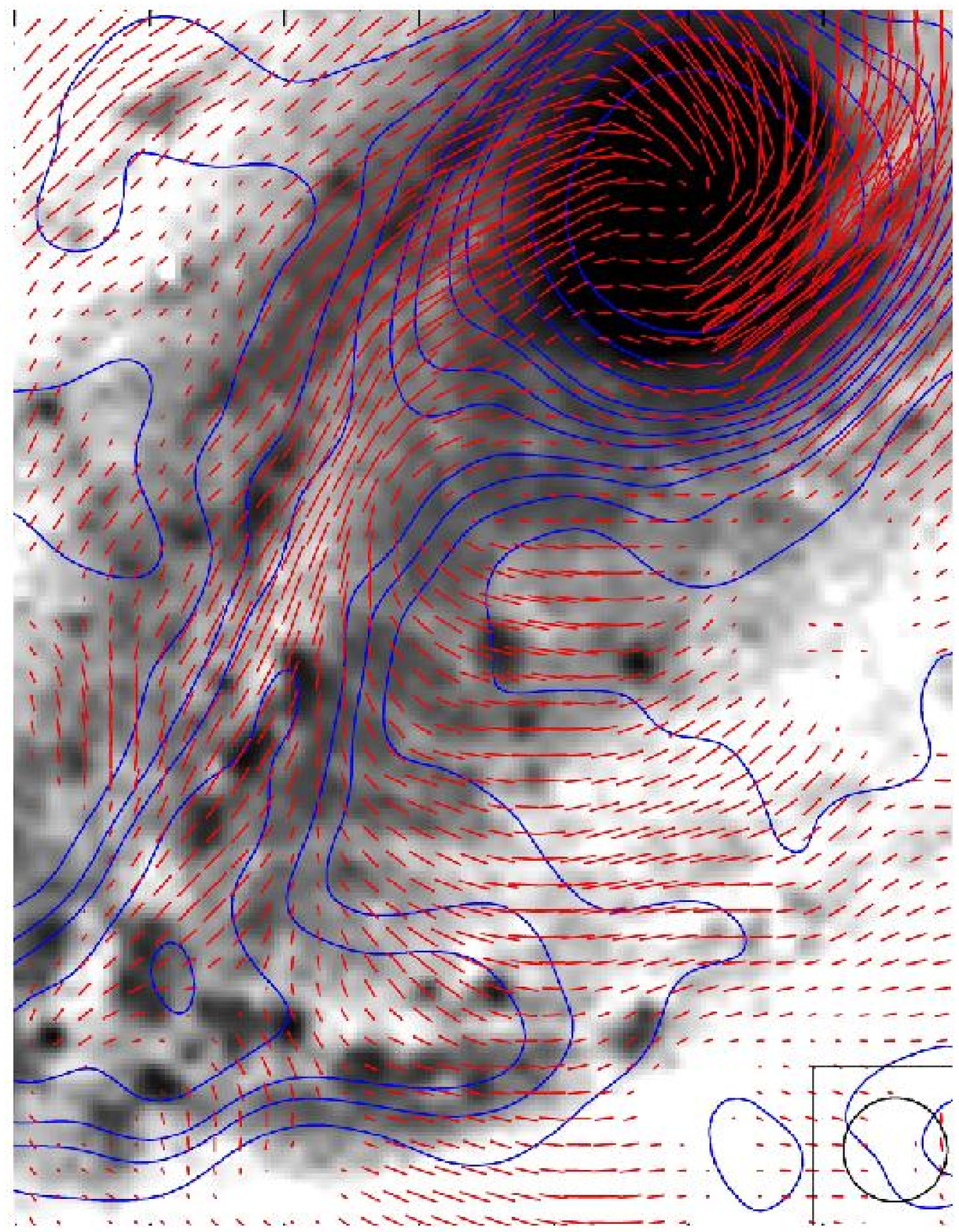}
\caption{Total radio emission (contours) and $B$--vectors of the
barred galaxy NGC~1097, observed at 6~cm wavelength with the VLA and
smoothed to 10'' resolution (\cite{beck+05}). The background optical
image is from Halton Arp (Copyright: MPIfR Bonn and Cerro Tololo
Observatory).} \label{fig:n1097}
\end{minipage}
\end{figure}

The intensity of {\em synchrotron emission}\ is a measure of the
number density of cosmic-ray electrons in the relevant energy range
and of the strength of the total magnetic field component in the sky
plane. Polarized emission emerges from ordered fields. As
polarization ``vectors'' are ambiguous by $180^\circ$, they cannot
distinguish {\em regular fields} with a constant direction within
the telescope beam from {\em anisotropic fields}\ which are
generated from turbulent magnetic fields by compressing or shearing
gas flows and frequently reverse their direction on small scales.
Unpolarized synchrotron emission indicates {\em turbulent fields}\
with random directions which have been tangled or generated by
turbulent gas flows.

The intrinsic degree of linear polarization of synchrotron emission
is about 75\%. The observed degree of polarization is smaller due to
the contribution of unpolarized thermal emission, which may dominate
in star-forming regions, by {\em Faraday depolarization}\ along the
line of sight and across the beam (\cite{sokoloff98}), and by
depolarization due to variations of the field orientation within the
beam and along the line of sight.

At short (e.g. centimeter) radio wavelengths the orientation of the
observed $B$--vector is generally parallel to the field orientation,
so that the magnetic patterns of many galaxies could be mapped
directly (\cite{beck05}). The orientation of the polarization
vectors is changed in a magnetized thermal plasma by {\em Faraday
rotation}. The rotation angle increases with the plasma density, the
strength of the component of the field along the line of sight and
the square of the observation wavelength. As the rotation angle is
sensitive to the sign of the field direction, only regular fields
can give rise to Faraday rotation, while anisotropic and random
fields do not. For typical plasma densities and regular field
strengths in the interstellar medium of galaxies, Faraday rotation
becomes significant at wavelengths larger than a few centimeters.
Measurements of the Faraday rotation from multi-wavelength
observations allow to determine the strength and direction of the
regular field component along the line of sight. Its combination
with the total intensity and the polarization vectors can yield the
three-dimensional picture of the magnetic field and allows to
distinguish the three field components: {\em regular, anisotropic
and random}.

\section{Total synchrotron emission: Tracer of star formation}

The integrated flux densities of total radio continuum emission at
centimeter wavelengths (frequencies of a few GHz), which is mostly
of nonthermal synchrotron origin, and far-infrared (FIR) emission of
star-forming galaxies are tightly correlated, first reported by
\cite{jong85}. This correlation achieved high popularity in galactic
research as it allows to use radio continuum emission as a
extinction-free tracer of star formation. The correlation holds for
starburst galaxies (\cite{lisenfeld96b}) as well as for blue compact
and low-surface brightness galaxies (\cite{chyzy07}). It extends
over five orders of magnitude (\cite{bell03}) and is valid to
redshifts of at least 3 (\cite{seymour08}), so that radio emission
serves as a star formation tracer in the early Universe. Only
galaxies with very recent starbursts reveal significantly smaller
radio-to-FIR ratios because the timescale for the acceleration of
cosmic rays and/or for the amplification of magnetic fields is
longer than that of dust heating (\cite{roussel03}).

Strongest total synchrotron emission (tracing the total, mostly
turbulent field) generally coincides with highest emission from dust
and gas in the spiral arms: The correlation also holds for the local
radio and FIR or mid-IR (MIR) intensities {\em within galaxies}
(e.g. \cite{beck88,hoernes98,walsh02,taba07a}). The highest
correlation of all spectral ranges is found between the total
intensity at $\lambda$6~cm and the mid-infrared dust emission, while
the correlation with the cold gas (traced by the CO(1-0) transition)
is less tight (\cite{frick01,walsh02,nieten06}). A wavelet
cross-correlation analysis for M~33 showed that the radio--FIR
correlation holds for all scales down to 1~kpc (\cite{taba07a}). The
correlation breaks down below scales of about 50~pc
(\cite{hughes06}) and in radio halos, probably due to the smoothing
effect on synchrotron intensity by cosmic-ray propagation.

If the thermal and nonthermal radio components are separated with
help of H$\alpha$ and FIR data (e.g. within M~33, \cite{taba07b}),
an almost perfect correlation is found between thermal radio and
infrared intensities at all scales. The nonthermal--FIR correlation
is less pronounced but highly significant. The polarized synchrotron
intensity, tracing the ordered field, is anticorrelated or not
correlated with all tracers of star formation (\cite{frick01}).

It is not obvious why the nonthermal synchrotron and the thermal FIR
intensities are so closely related. The intensity of synchrotron
emission depends not only on the density of cosmic-ray electrons
(CRE), but also on about the square of the strength of the total
magnetic field $B_\mathrm{t}$ (its component in the sky plane, to be
precise). The radio--FIR correlation requires that magnetic fields
and star-formation processes are connected. If $B_\mathrm{t}$ is
strong, most of the cosmic-ray energy is released via synchrotron
emission within the galaxy, the CRE density decreases with
$B_\mathrm{t}^2$ and the integrated radio synchrotron luminosity
depends on the CRE injection rate, not on $B_\mathrm{t}$. If most
thermal energy from star formation is also emitted within a galaxy
via far-infrared emission by warm dust, this galaxy can be treated
as a ``calorimeter'' for thermal and nonthermal emission. Prime
candidates for ``calorimeter'' galaxies are those with a high
star-formation rate (SFR). If $B_\mathrm{t}$ increases with SFR
according to $B_\mathrm{t}\propto SFR^{0.5}$, a linear radio--FIR
correlation for the integrated luminosities is obtained
(\cite{lisenfeld96a},b). However, the calorimeter model cannot
explain the local correlation within galaxies.

In galaxies with low or moderate SFR and $B_\mathrm{t}$, synchrotron
lifetime of CRE is sufficiently large to leave the galaxy. To obtain
a global or local radio--FIR correlation, coupling of magnetic
fields to the gas clouds is needed. A scaling
$B_\mathrm{t}\propto\rho^{1/2}$ was proposed
(\cite{helou93,niklas97,hoernes98}) where $\rho$ is the average
density of the neutral gas. A nonlinear correlation (with a slope of
about 1.3) between the nonthermal radio luminosity and the FIR
luminosity from warm dust is achieved by further assuming energy
equipartition between magnetic fields and cosmic rays and a Schmidt
law of star formation ($SFR\propto\rho^{1.5}$) (\cite{niklas97}). In
this model the total magnetic field strength and the star-formation
rate $SFR$ are related via $B_\mathrm{t}\propto SFR^{0.3}$.

The radio--FIR correlation indicates that {\em equipartition}\
between the energy densities of the total magnetic field and the
total cosmic rays is valid, at least on spatial scales larger than
about 100~pc (Stepanov et al., this volume) and on timescales of
larger than the CRE acceleration time (a few $10^6$ years). Then the
strength of the total magnetic field can be determined from the
intensity of the total synchrotron emission, assuming a ratio $K$
between the numbers of cosmic-ray protons and electrons in the
relevant energy range (usually $K\simeq100$). In regions where
electrons lost already a significant fraction of their energy, e.g.
in strong magnetic fields or radiation fields or far away from their
places of origin, $K$ is $>100$ and the standard value of 100 yields
an underestimate (\cite{beck+krause05}). On the other hand, in case
of field fluctuations along the line of sight or across the
telescope beam, the equipartition value is an overestimate
(\cite{beck03}).

The typical average equipartition strength of the total magnetic
field in spiral galaxies is about $10~\mu$G. Radio-faint galaxies
like M~31 and M~33, our Milky Way's neighbors, have weaker total
magnetic fields (about $5~\mu$G), while gas-rich galaxies with high
star-formation rates, like M~51 (Fig.~\ref{fig:m51}), M~83 and
NGC~6946, have total field strengths of 20--30~$\mu$G in their
spiral arms. The degree of radio polarization within the spiral arms
is only a few \%; hence the field in the spiral arms must be mostly
tangled or randomly oriented within the telescope beam, which
typically corresponds to a few 100~pc. Turbulent fields in spiral
arms are probably generated by turbulent gas motions related to
supernovae (\cite{avillez05}), stellar winds, spiral shocks
(\cite{dobbs08}) or a small-scale turbulent dynamo (\cite{beck96}).

The mean energy densities of the magnetic field and of the cosmic
rays in NGC~6946 and M~33 are $\simeq10^{-11}$~erg~cm$^{-3}$ and
$\simeq10^{-12}$~erg~cm$^{-3}$, respectively (\cite{beck07,taba08}
and this volume), about 10 times larger than that of the ionized
gas, but similar to that of the turbulent gas motions across the
whole star-forming disk. The magnetic energy possibly dominates in
the outer disk of NGC~6946.

The strongest total fields of 50--100~$\mu$G are found in starburst
galaxies, like M~82 (\cite{klein88}) and the ``Antennae'' NGC~4038/9
(\cite{chyzy04}), and in nuclear starburst regions, like in the
centers of NGC~1097 and other barred galaxies (\cite{beck+05}). In
starburst galaxies, however, the equipartition field strength per
average gas surface density is much lower than in normal spirals.
This indicates strong energy losses of the cosmic-ray electrons, so
that the equipartition field strength is probably underestimated by
a factor of a few (\cite{thompson06}). This was recently confirmed
by Zeeman measurements of OH maser lines (\cite{robishaw08}).


In case of energy equipartition, the scale length of the total field
in the disk of galaxies is at least $(3-\alpha)$ times larger than
the synchrotron scale length of typically 4~kpc (where
$\alpha\simeq$--1 is the synchrotron spectral index). The resulting
value of $\simeq$16~kpc is a lower limit because the CRE lose their
energy with distance from the star-forming disk and the
equipartition assumption yields too small values for the field
strength. The galactic fields extend far out into intergalactic
space. The same argument holds for the vertical extent of radio
halos around galaxies which is also limited by CRE energy losses.
The dumbbell-shaped halo around edge-on galaxies like NGC~253
(\cite{heesen08} and this volume) is the result of enhanced
synchrotron and Inverse Compton losses in the inner region.


As proposed by \cite{battaner00}, the magnetic field energy density
may generally reach the level of global rotational gas motion and
affect galaxy rotation in the outermost parts of spiral galaxies. At
GHz frequencies the measured extent of the radio disks of galaxies
is limited by energy loss of cosmic-ray electrons, so that
measurements at low frequencies (where energy losses are smaller)
are needed, e.g. with LOFAR (\cite{beck08}). Faraday rotation
towards polarized background sources may allow to measure weak
fields to even larger distances from the star-forming disks.

\section{Polarized synchrotron emission: Tracer of ordered fields}

\begin{figure}
\parbox[b]{8cm}{
\includegraphics[bb = 47 213 522 573,width=9cm,clip=]{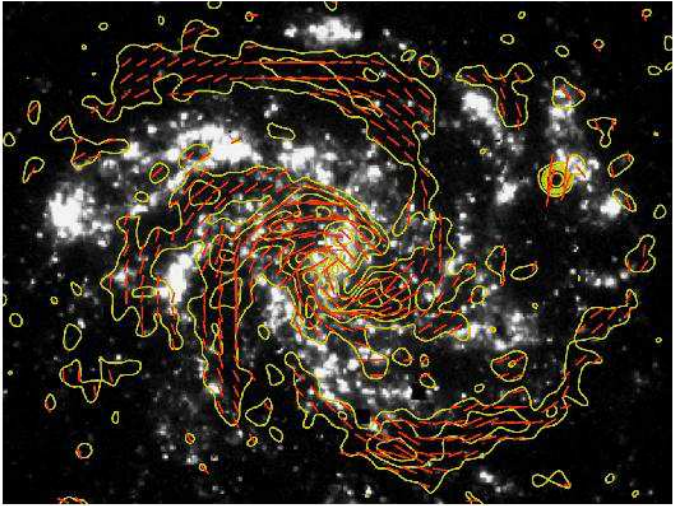} }
\hfill
\parbox[b]{4cm}{
\caption{Polarized radio emission (contours) and $B$--vectors of
NGC~6946 (15'' resolution), combined from observations at 6~cm
wavelength with the VLA and Effelsberg 100m telescopes
(\cite{beck+hoernes96}). The background image shows the H$\alpha$
emission (\cite{ferguson98}) (Copyright: MPIfR Bonn. Graphics:
\textit{Sterne und Weltraum}). } \label{fig:n6946} }
\end{figure}

The ordered (regular and/or anisotropic) fields traced by the
polarized synchrotron emission are generally strongest
(10--15~$\mu$G) in the regions {\em between}\ the optical spiral
arms, oriented parallel to the adjacent optical spiral arms. In some
galaxies the field forms {\em magnetic arms}\ between the optical
arms, like in NGC~6946 (Fig.~\ref{fig:n6946}). These are probably
generated by a large-scale dynamo. In galaxies with strong density
waves some of the ordered field is concentrated at the inner edge of
the spiral arms, e.g. in M~51 (\cite{patrikeev06}), but the
arm--interarm contrast of the ordered field is small, much smaller
than that of the random field.

The ordered magnetic field forms spiral patterns in almost every
galaxy (\cite{beck05}), even in ring galaxies (\cite{chyzy08}) and
in flocculent galaxies without an optical spiral structure
(\cite{soida02}). Hence, the field lines generally do {\em not}\
follow the (almost circular) gas flow and need dynamo action to
obtain the required radial field components. Spiral fields with
large pitch angles are also observed in the central regions of
galaxies and in circum-nuclear gas rings (\cite{beck+05}).

\begin{figure}[htb]
\begin{center}
\includegraphics[bb = 47 41 522 245,width=13cm,clip=]{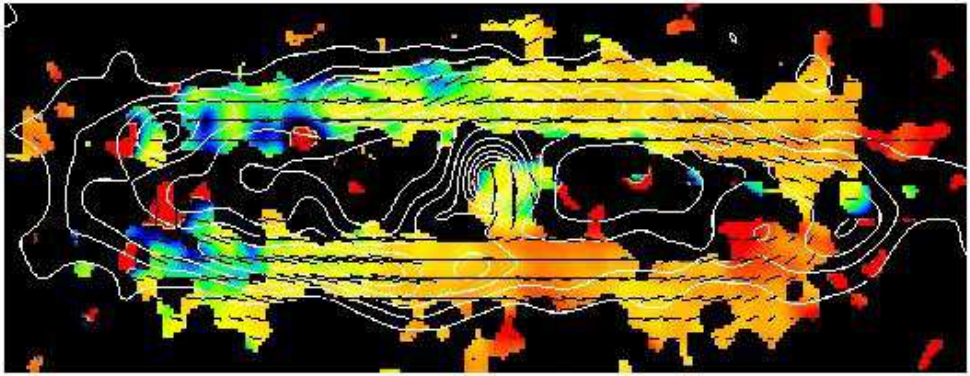}
\caption{Total radio emission (contours) and $B$--vectors at 6~cm
and Faraday rotation measures (RM) between 6~cm and 11~cm (colors)
of the Andromeda galaxy M~31 (5' resolution), observed with the
Effelsberg telescope. The color scale ranges from RM = -80 to +120
rad$^{-2}$ (\cite{berk03}) (Copyright: MPIfR Bonn).} \label{fig:m31}
\end{center}
\end{figure}

In galaxies with massive bars the field lines follow the gas flow
(Fig.~\ref{fig:n1097}). As the gas rotates faster than the bar
pattern of a galaxy, a shock occurs in the cold gas which has a
small sound speed, while the flow of warm, diffuse gas is only
slightly compressed but sheared. The ordered field is also hardly
compressed. It is probably coupled to the diffuse gas and strong
enough to affect its flow (\cite{beck+05}, Fletcher et al., this
volume). The polarization pattern in spiral arms and bars can be
used as a tracer of shearing gas flows in the sky plane and hence
complements spectroscopic measurements.

Nearby galaxies seen edge-on generally show a disk-parallel field
near the disk plane (\cite{dumke95}), so that polarized emission can
also be detected from distant, unresolved galaxies (\cite{stil08}).
High-sensitivity observations of edge-on galaxies like NGC~891
(\cite{krause08}) and NGC~253 (Heesen et al., this volume) revealed
vertical field components in the halo forming an X-shaped pattern
(Krause, this volume). This is inconsistent with the predictions
from standard dynamo models. The field is probably transported from
the disk into the halo by an outflow emerging from the disk. The
similarity of scale heights of radio halos in different galaxies
indicates that the outflow velocity increases with the
star-formation rate (Krause, this volume). Interestingly, a recent
model for global outflows from galaxy disks (neglecting magnetic
fields) shows an X-shaped velocity field (\cite{vecchia08}).
Improved outflow models including magnetic fields and dynamo action
are needed.

\section{Faraday rotation: Tracer of regular dynamo fields}

Faraday rotation is a signature of regular (coherent) fields which
could be generated by the mean-field (or large-scale) {\em dynamo}
(\cite{elstner92,beck96}; Elstner, this volume). Dynamo fields are
described by modes of different azimuthal symmetry in the disk plane
and vertical symmetry (even or odd parity) perpendicular to the disk
plane. Several modes can be excited in the same object. In flat,
rotating objects like galaxy disks, the strongest mode S0 consists
of a toroidal field of {\em axisymmetric spiral}\ shape within the
disk, without sign reversals across the equatorial plane, and a
weaker poloidal field of quadrupolar structure with a reversal of
the vertical field component across the plane. Antisymmetric
(A--type) fields are generated preferably in spherical objects like
halos (\cite{moss08}). The magneto-rotational instability (MRI)
supports symmetric fields while primordial seed fields support
antisymmetric fields (\cite{krause+beck98}).


Spiral dynamo modes can be identified from the pattern of
polarization angles and Faraday rotation measures (RM) from
multi-wavelength radio observations of galaxy disks
(\cite{krause90,elstner92}) or from RM data of polarized background
sources (\cite{stepanov08}). The disks of a few spiral galaxies
indeed reveal large-scale RM patterns, as predicted. The Andromeda
galaxy M~31 hosts a dominating axisymmetric disk field (mode S0)
(Fig.~\ref{fig:m31} and \cite{fletcher04}). Other candidates for a
dominating axisymmetric disk field are the nearby spiral IC~342
(\cite{krause89}) and the irregular Large Magellanic Cloud (LMC)
(\cite{gaensler05}). The magnetic arms in NGC~6946 can be described
by a superposition of two azimuthal dynamo modes which are phase
shifted with respect to the optical arms (\cite{beck07}). However,
in many observed galaxy disks no clear patterns of Faraday rotation
were found. Either several dynamo modes are superimposed and cannot
be distinguished with the limited sensitivity and resolution of
present-day telescopes, or the timescale for the generation of
large-scale modes is longer than the galaxy's lifetime. Dynamo
models predict a rapid amplification of small-scale fields already
in protogalaxies, while the generation of fully coherent large-scale
fields takes several Gyrs, depending on the galaxy size
(\cite{arshakian08} and this volume). Furthermore, anisotropic
fields dominate over dynamo modes if shearing gas flows are strong
(see Sect.~4).

Vertical fields as predicted by dynamo models should generate
large-scale RM patterns around edge-on galaxies, but so far
indication for an antisymmetric poloidal field was found only in the
halo of NGC~253 (Heesen et al., this volume). Previous indirect
evidence for symmetric fields from the dominance of inward-directed
fields (\cite{krause+beck98}) is no longer supported by the
increased sample of galaxies.

Faraday rotation in the direction of QSOs allows to determine the
field pattern in an {\em intervening galaxy} (\cite{stepanov08}).
This method can be applied to much larger distances than the
analysis of RM of the polarized emission from the foreground galaxy
itself. Faraday rotation of QSO emission in distant, intervening
galaxies revealed significant regular fields of several $\mu$G
strength (\cite{bernet08,kronberg08}) which is a challenge for
classical dynamo models (\cite{arshakian08}).

The recently developed method of {\em RM Synthesis}, based on
multi-channel spectro-polarimetry, transforms the spectral data cube
into a data cube of maps in Faraday depth space (\cite{brentjens05},
Heald, this volume). RM components from distinct regions along the
line of sight can be distinguished by their positions and widths in
Faraday depth. Faraday screens appear as sharp lines, while emitting
and rotating layers form broad structures in Faraday depth. The
transformation of Faraday depth into geometrical depth ({\em Faraday
tomography}) needs modeling.

\section{Magnetic fields in the Milky Way}

Surveys of the total synchrotron emission from the Milky Way yield
equipartition strengths of the total field of 6~$\mu$G near the Sun
and about 10~$\mu$G in the inner Galaxy (Berkhuijsen, in
\cite{wielebinski05}), consistent with Zeeman splitting data of
low-density gas clouds (Heiles, this volume). In the nonthermal
filaments near the Galactic center the field strength may reach
several 100~$\mu$G (\cite{reich94}). Milligauss fields were found in
pulsar wind nebulae from the break in the synchrotron spectrum
(e.g., \cite{kothes08}).

The observed degree of radio and optical polarization in the local
Galaxy implies a ratio of ordered to total field strengths of
$\simeq0.6$ (\cite{heiles96}). For a total field of 6~$\mu$G this
gives 4~$\mu$G for the local ordered field component (including
anisotropic fields). Faraday RM and dispersion measure data of
pulsars give an average strength of the local coherent regular field
of $1.4\pm0.2~\mu$G (\cite{rand94}). In the inner Norma arm, the
average strength of the coherent regular field is $4.4\pm0.9~\mu$G
(\cite{han02}).

The all-sky maps of polarized synchrotron emission at 1.4~GHz from
the Milky Way from DRAO and Villa Elisa and at 22.8~GHz from WMAP,
and the new Effelsberg RM survey of polarized extragalactic sources
were used to model the large-scale Galactic field (\cite{sun08}).
One large-scale reversal is required about 1--2~kpc inside the solar
radius, which also agrees with the detailed study of RMs from
extragalactic sources near the Galactic plane (\cite{brown07},
Kothes \& Brown, this volume). Recent pulsar RM values indicate
reversals at several Galactic radii (Han, this volume). None of the
models for a simple large-scale field structure of the Milky Way
survived a statistical test (\cite{men08}). Similar to external
galaxies, the Milky Way's regular field probably has a complex
structure which can be revealed only based on a larger sample of
pulsar and extragalactic RM data.

\begin{figure}
\parbox[b]{8cm}{
\includegraphics[bb = 20 20 592 373,width=9cm,clip=]{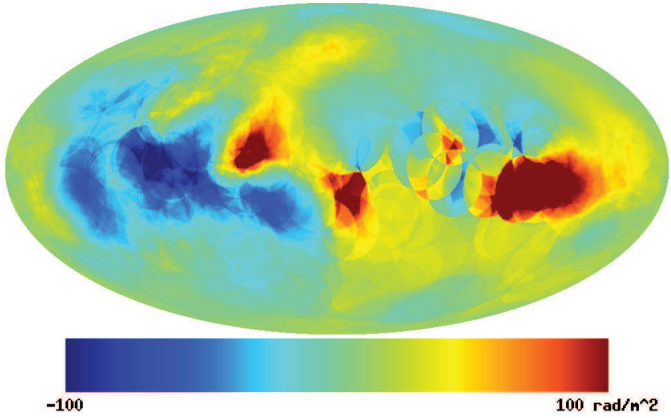} }
\hfill
\parbox[b]{4cm}{
\caption{All-sky map of rotation measures in the Milky Way,
generated from data towards about 800 polarized extragalactic
sources compiled by \cite{johnston04} and from about 1800 data of
the new Effelsberg RM survey (from \cite{sun08}).} \label{fig:sun} }
\end{figure}

RMs of extragalactic sources and of pulsars reveal no large-scale
reversal across the plane for Galactic longitudes
l=90$^\circ$--270$^\circ$ (Fig.~\ref{fig:sun}): the local disk field
is part of a large-scale symmetric (quadrupolar) field structure.
However, towards the inner Galaxy (l=270$^\circ$--90$^\circ$) the RM
signs are {\em opposite}\ above and below the plane
(Fig.~\ref{fig:sun}). This may indicate a global antisymmetric mode
in the halo (\cite{han97}), similar to the results for the spiral
galaxy NGC~253 (Heesen et al., this volume). Note that a
superposition of a disk field with even parity and a halo field with
odd parity cannot be explained by classical dynamo theory
(\cite{moss08}).

While the large-scale field is much more difficult to measure in the
Milky Way than in external galaxies, Galactic observations can trace
magnetic structures to much smaller scales (\cite{wielebinski05},
Reich \& Reich, this volume).  Small-scale and turbulent field
structures are enhanced in spiral arms compared to interarm regions,
as indicated by the different slopes of their {\em RM structure
functions} (\cite{haverkorn06}). The all-sky and the Galactic plane
polarization surveys at 1.4~GHz (Reich \& Reich and Landecker et
al., this volume) reveal a wealth of structures on pc and sub-pc
scales: filaments, canals, lenses and rings. Their common property
is to appear only in the maps of polarized intensity, but not in
total intensity. Some of these are artifacts due to strong
depolarization of background emission in a foreground Faraday
screen, called {\em Faraday ghosts}, but carry valuable information
about the turbulent ISM (\cite{fletcher06}). Faraday rotation in
foreground objects (e.g. supernova remnants, planetary nebulae,
pulsar wind nebulae or photo-dissociation regions of molecular
clouds) embedded in diffuse polarized Galactic emission may generate
a {\em Faraday shadow}\ or {\em Faraday screen}\ which enable to
estimate the regular field strength by modeling the radiation
transfer (\cite{wolleben04,ransom08}, Reich \& Reich, this volume).

\section{Outlook}

Future radio telescopes will widen the range of observable magnetic
phenomena. High-resolution, future deep observations at high
frequencies with the Extended Very Large Array (EVLA) and the Square
Kilometre Array (SKA) (Gaensler, this volume) will directly show the
detailed field structure and the interaction with the gas. The SKA
will also measure the Zeeman effect in much weaker magnetic fields
in the Milky Way and nearby galaxies. Forthcoming low-frequency
radio telescopes like the Low Frequency Array (LOFAR) and the
Murchison Widefield Array (MWA) will be suitable instruments to
search for extended synchrotron radiation at the lowest possible
levels in outer galaxy disks, halos and clusters, and the transition
to intergalactic space. At low frequencies we will get access to the
so far unexplored domain of weak magnetic fields in galaxy halos
(\cite{beck08}). The detection of radio emission from the
intergalactic medium will allow to probe the existence of magnetic
fields in such rarified regions, measure their intensity, and
investigate their origin and their relation to the structure
formation in the early Universe. Low frequencies are also ideal to
search for small Faraday rotation from weak interstellar and
intergalactic fields. ``Cosmic magnetism'' is the title of Key
Science Projects for LOFAR and SKA.



\end{document}